\documentclass[11pt]{article}

\usepackage{aaspp4,eqsecnum,astrobib,graphicx,rotating}

\begin{document}

\epsfverbosetrue

\begin{flushright}
\today
\end{flushright}

\vglue 50 mm

\title{Are Seyfert Narrow Line Regions Powered by Radio Jets?}

\author{Geoffrey V. Bicknell \altaffilmark{1}, 
Michael A. Dopita \altaffilmark{2},
Zlatan I. Tsvetanov \altaffilmark{3},
Ralph S. Sutherland \altaffilmark{1}
}

\vglue 50 mm

\begin{center}
Submitted to the Astrophysical Journal, Sept. 4, 1997
\end{center}

\altaffiltext{1}{ANU Astrophysical Theory Centre, Australian National University, Canberra, ACT 0200,
Australia. The ANUATC is operated jointly by the  Mt. Stromlo and Siding Spring Observatories and the
School of Mathematical Sciences of the Australian National University.}
\altaffiltext{2}{Mount Stromlo \& Siding Spring Observatories, Weston PO, ACT 2611, Australia}
\altaffiltext{3}{Dept. of Physics \& Astronomy, Johns Hopkins University, Baltimore, MD  21218, USA}

\newpage

\begin{abstract}

We argue that the narrow line regions of Seyfert galaxies are powered by the transport of energy and
momentum by the radio-emitting jets. This implies that the ratio of the radio power to jet energy flux is
much smaller than is usually assumed for radio galaxies. This can be partially attributed to the smaller
ages ($\sim 10^6 \> \rm yrs$) of Seyferts compared to radio galaxies but one also requires that either the
magnetic energy density is more than an order of magnitude below the equipartition value, or more likely,
that the internal energy densities of Seyfert jets are dominated by thermal plasma, as distinct from the
situation in radio galaxy jets where the jet plasma is generally taken to be nonthermally dominated. If
the internal energy densities of Seyfert jets are initially dominated by relativistic plasma, then an
analysis of the data on jets in five Seyfert galaxies shows that all but one of these would have mildly
relativistic jet velocities near 100~pc in order to power the respective narrow-line regions. However,
observations of jet-cloud interactions in the NLR provide additional information on jet velocities and
composition via the momentum budget. Our analysis of a jet-cloud interaction in NGC~1068, 24~pc from the
core, implies a shocked jet pressure much larger than the minimum pressure of the radio knot, a velocity
(probably accurate to within a factor of a few) $\sim 0.06 \, c$ ($18,000 \> \rm km
\> s^{-1}$) and a temperature of thermal gas in the jet
$\sim 10^9 \> \rm K$ implying mildly relativistic electrons but thermal protons.  The estimated jet velocity
is proportional to the jet energy flux and provides an independent argument that the energy flux in the
northern NGC~1068 jet is much greater than previously supposed and is capable of providing
significant energy input to the narrow line region. The jet mass flux at this point $\sim  0.5 M_\odot \>
yr^{-1}$, is an order of magnitude higher than the mass accretion rate $\sim 0.05 \> M_\odot \> yr^{-1}$
estimated from the bolometric luminosity of the nucleus, strongly indicating entrainment into the jet and
accompanying deceleration. Consequently, the jet velocity near the black hole is likely to be mildly
relativistic. We estimate an initial jet mass flux $\sim 0.02 \> M_\odot \> \rm yr^{-1}$ which is comparable
to the mass accretion rate. This mass flux is consistent with the densities inferred for accretion disc
coronae from high energy observations, together with an initially mildly relativistic velocity and an
initial jet radius of order 10 gravitational radii.

\end{abstract}

\newpage

\section{Introduction}

The dominant paradigm for the excitation of emission lines in the narrow line regions (NLRs)  of
Seyfert galaxies invokes photoionization by a central power-law continuum. The apparent lack of
ionizing photons in Seyfert~2s has been attributed to obscuration by a thick torus surrounding the
central region of the Active Galactic Nucleus (AGN). Nevertheless,
\citeN{binette93a} have argued that there is a deficit of ionizing photons in {\em all} AGN
leading to some scepticism of central power-law photoionization.  On the other hand, it has been
recognized for some time \cite{binette85} that excitation by autoionizing shocks in various classes of
AGN can give a reasonable fit to the observed spectra and improve the agreement between theory
and observation for the temperature-sensitive [OIII]$\lambda4363$ line. Notable individual successes
of this approach include a self-consistent model for the excitation of the emission line filaments in
Centaurus A
\cite{sutherland93b} and the excitation of the disk in the central 100~pc of M87 \cite{dopita97a}.  In
the latter case,  detection by HST of high excitation lines in the ultraviolet together with the
excellent agreement between the observed and predicted spectra make a strong case for shock excitation.
\citeN{bicknell97a} have also shown that the narrow-line emission from Gigahertz Peak Spectrum (GPS)
and Compact Steep  Spectrum (CSS) radio sources can plausibly be related to the excitation of a dense
ISM by the bow shock surrounding the radio source. That model has the additional feature of providing
a natural explanation for the peak in the radio spectrum of these sources. \citeN{koekemoer97a} have
also made a strong case for the shock excitation of the extended emission-line region in the radio
galaxy PKS2356-61.  For Seyfert galaxies,
\citeN{viegas97a} have also suggested that emission lines from shocked gas in cloud-cloud collisions
constitute  a fraction of the total emission line output from the narrow line region (NLR). In view of
the increasing interest in autoionizing shocks,
a grid of models of shock excited spectra has been calculated \cite{dopita96a,dopita96b}. These provide a
good description of the observed spectra.

The correlation between radio power and [OIII]$\lambda5007$ luminosity in Seyfert galaxies discovered
by \citeN{whittle85} over twelve years ago, suggests an intimate connection between  the
radio-emitting and emission-line components. Indeed, where
observations are made at high enough resolution, detailed correspondence between radio and optical features
is often found, along with evidence of dynamical structure and sources of local ionization indicative of
shocks (\shortciteN{whittle88a},\citeN{goodrich92a,capetti96a}). In the specific case of NGC~1068
\citeN{capetti97a} have shown, using HST/WFPC2 [OIII] and H$\alpha$+[NII] images overlaid on a VLA radio
image of the radio source, that there is a close correspondence between the radio and optical structure,
strongly supporting the view that the NLR of NGC~1068 is the result of interaction with the radio
plasma.  Another key observation is that the radio axes of Seyferts correspond remarkably to the axes of
their ionization cones, again suggesting that the radio and optical emission are causally related
\cite{wilson94a}. The view advocated here is that the Whittle radio--optical luminosity correlation
together with the unequivocal morphological correlations between radio and line-emitting gas is
strong circumstantial evidence that the  mechanical energy and momentum transported by the radio
plasma is responsible for the excitation of at least some, and probably most, of the narrow-line
region (NLR).  An alternative interpretation of a similar correlation for radio galaxies was given by
\citeN{rawlings91} who suggested  that the accretion disk luminosity  (as indicated by the NLR [OIII]
luminosity)  and the mechanical luminosity of the jet are correlated. This suggestion was taken up by
\citeN{falcke95a} and \citeN{falcke95b} who proposed that a substantial fraction  of the energy
dissipated by all AGN accretion disks finds its way into jet mechanical luminosity. However, this
view does not really represent an alternative, since the inferred jet energy flux is 
comparable to the ionizing luminosity of the accretion disk and capable of having an important
influence on the energy budget of the NLR gas.

If the narrow line emission from Seyfert galaxies is to be powered by the expansion of the radio
lobe, then, appealing to the GPS/CSS model of \citeN{bicknell97a} as a basis for order of magnitude
estimates, the [OIII]$\lambda5007$ luminosity is given by
\begin{equation}
L([OIII]) \approx  1.0 \times 10^{-2} \> 
\left( \frac {6}{8-\delta} \right) \: F_{\rm E} \> {\rm ergs \> s^{-1}} 
\end{equation}
where $\delta$ is the power-law index of the ambient density distribution and $F_{\rm E}$ is the
combined energy flux in both jets. Hence, a median Seyfert [OIII] luminosity $\sim 10^{42} \: 
\rm ergs \> s^{-1}$ requires an energy flux in each jet $\sim 5 \times 10^{43} \> \rm ergs \> s^{-1}$.
Thus, a major argument {\em against} the nonthermal plasma powering the emission-line region (e.g.. 
\citeN{wilson97a}) is based upon the low radio luminosity of Seyferts (typically $\sim 10^{23} \> 
\rm W \: Hz^{-1}$ at 5~GHz). Using standard
 conversion factors relating radio luminosity and mechanical jet power used for radio galaxies,  one
finds that the jet power falls short of what is required by at least two orders of magnitude.  In
earlier papers (e.g.. \citeN{bicknell95a}, \citeN{bicknell97a}), we have represented this conversion
factor by $\kappa_\nu$, the ratio of monochromatic power at frequency $\nu$ to jet energy flux. The
advantage of using this parameter is that it is easily  estimated using straightforward physics. The 
correlation between  [OIII]$\lambda5007$ luminosity and 1.4~GHz radio power is represented in
Figure~\ref{f:oiii} where the  correlation is plotted for a number of radio loud objects (used  by
\citeN{bicknell97a}) together with
the Seyferts in Whittle's sample. The lines represent the theoretical relationship between [OIII]
luminosity and radio power predicted by the GPS/CSS model of \citeN{bicknell97a}. The values of
$\log \kappa_{1.4}$ (where $\kappa_{1.4}$ represents $\kappa_\nu$ at a frequency of $1.4 \> \rm GHz$)
range from -14 (extreme left)  to -10 (extreme right).  Seyferts are separated quite clearly from the
radio loud  objects by, on average, three or even four orders of magnitude in $\kappa_{1.4}$

In this paper, we examine the question of whether Seyferts are in fact characterized by such
low values of $\kappa_{1.4}$ and additional aspects of the energy and momentum budgets which have a
strong bearing on the question of whether jets can power the NLR of Seyfert galaxies. Some of the following
analysis is specific to NGC~1068 and the case of this galaxy, we examine the
dyanmical link between the jet parameters and the expected parameters of the corona above the accretion disk
in the vicinity of the black hole.

\newpage

\section {The Relation between Jet Power and Radio Luminosity.}

Traditionally, an effective value of $\kappa_\nu$ has been based on an assumed ``efficiency of conversion''
of jet power into total radio luminosity. However,
 this ``efficiency factor'' has a physical basis (\citeN{bicknell84b}, \citeN{eilek89},
\citeN{bicknell95a}, \citeN{begelman96a},
\citeN{bicknell97a}). For convenience, we summarize here the physics of this relationship as given by
\citeN{bicknell97a}. Consider a jet-fed radio lobe: The total energy, $E_{\rm L}$ accumulating in the lobe
can be represented approximately as 
$f_{\rm ad} \, F_{\rm E} \, t$ where $F_{\rm E}$ is the jet energy flux, $t$ is the age of the lobe  and
$f_{\rm ad} \sim 0.5$ is a factor which accounts for the energy lost adiabatically by
expansion\footnote{$f_{\rm ad} =1$ if there is no energy lost by expansion.}. The synchrotron emission from
the lobe is given by
$P_\nu \approx C_{\rm syn}(a) \, K \, B^{(a+1)/2} \, \nu^{-(a-1)/2}, V$ where $C_{\rm syn}$ is expressed in
terms of the \citeN{pacholczyck70} synchrotron parameters, $c_i$ by
$C_{\rm syn}(a) = 4 \pi c_5(a) c_9(a) (2c_1)^{(a-1)/2}$, the electron energy density per unit energy,
$N(E)$, is given by $N(E) = KE^{-a}$,
$B$ is the  (randomly oriented) magnetic field, $\nu$ is the observing frequency, and $V$ is the volume. The
parameter
$K$ can be  expressed in terms of the electron/positron energy density, $\epsilon_{\rm e}$  via $K = (a-2)
\epsilon_{\rm e}  (\gamma_0 m_{\rm e} c^2)^{(a-2)} [1-(\gamma_1/\gamma_0)^{-(a-2)}]^{-1}$ where
$\gamma_0$ and
$\gamma_1$ are respectively the lower and upper Lorentz factor cutoffs in the electron distribution. 
Writing the  total electron/positron energy in the lobe as $f_{\rm e} E_{\rm L}$ where $f_{\rm e}$ is the 
electron/positron fraction of the total energy, we obtain for the ratio of monochromatic radio power to jet
energy flux:
\begin{equation}
\kappa_{\nu} \approx (a-2) \, C_{\rm syn}(a) \: (\gamma_0 m_{\rm e} c^2)^{(a-2)} \:
\left[ 1-(\gamma_1/\gamma_0)^{-(a-2)} \right]^{-1} \:  f_{\rm e} \, f_{\rm ad} \: B^{(a+1)/2} \: t
\end{equation} 
The most important factors in this equation are (1) the electron/positron fraction, $f_{\rm e}$,  (2) the
magnetic field, $B$ and (3) the age, $t$. This relationship is plotted in Figure~\ref{f:kappa_nu} for a
range of values of the evolutionary parameter $\tau = f_{\rm e} \, f_{\rm ad}
\, t$, assuming $\gamma_0=1$,
$\gamma_1= \infty$, a spectral index of $0.7$ and an observing frequency of $1.4 \: \rm GHz$. The band of
$\kappa_{1.4}$ ($-15 \la \log \kappa_{1.4} \la -13$) which we suggest is relevant to Seyferts is indicated.
For example a value of $\kappa_{1.4} \sim 10^{-13}$ and radio power $\sim 10^{23} \: W \> Hz^{-1}$ indicates
a combined jet energy flux  $\sim 10^{43} \: \rm ergs \: s^{-1}$. From the plot it is evident that the
values of $B$ and $\tau$ which are most conducive to $\kappa_{1.4}$ lying in this range are in the vicinity
of $B \sim 10^{-5} - 10^{-4} \> \rm G$ and corresponding values of 
$\tau \sim 10^6 - 10^4 \: \rm yrs$. Typical equipartition magnetic fields and dynamical ages for Seyfert
galaxies are approximately $10^{-4} \: \rm G$ and $10^6 \: \rm yrs$ \cite{wilson97a}, respectively, so that
values of $B$ and $\tau$ in the required range are feasible. However, if $\tau \sim 10^6 \> \rm yrs$ the
magnetic field is required to be approximately an order of magnitude below equipartition. If the magnetic
field is at equipartition we require $\tau \sim 10^4 \> \rm yrs$. The latter would be the case if $\tau$  is
less than the dynamical age as a result of  dominance of the thermal pressure (compared to the nonthermal)
making $f_{\rm e} \la 0.1$. Thus this analysis raises the prospect that Seyfert jets, on the kpc scale, are
dominated by thermal gas as distinct from radio galaxy and quasar jets which are normally taken to be
dominated by  nonthermal plasma. 

A related way of looking at this question is to ask whether the typical  equipartition pressures inferred
for the lobes of Seyfert galaxies are capable of driving the
$\sim 500 \> \rm km \> s^{-1}$ shocks necessary to explain the optical and UV emission line spectrum.  The
pressure  required to drive a shock of velocity $V_{\rm sh}$ into gas with pre-shock density $\rho_{\rm ps}$
is
$\rho_{\rm ps} V_{\rm sh}^2 \approx 6 \times 10^{-8} \, (n_{\rm ps}/10 \> \rm cm^{-3}) \, (V_{\rm sh} / 500
\> \rm km \> s^{-1})^2$ where $n_{\rm ps}$ is the pre-shock Hydrogen density. In order to produce the
typical [SII] densities $\sim 10^{3-4} \> \rm cm^{-3}$ inferred for NLR clouds we  require $n_{\rm ps} \ga
10 \> cm^{-3}$ requiring driving pressures 
$\ga 10^{-7} \> \rm dyn \> cm^{-2}$ well above the typical values of 
$\sim 10^{-9} \> \rm dyn \> cm^{-2}$ estimated for the lobes of typical Seyfert~2s from the nonthermal
emission. Hence, we are again forced to the conclusion that either the lobes are either well out of
equipartition or else that the lobe plasma contains a significant thermal component.

\newpage

\section{Implications for the ISM of Seyfert Galaxies}

Typically Seyfert radio sources are of the order of approximately a kpc in size. This statistic,
combined with our inference of the energy flux in the jets required to power the narrow line region,
has implications for the density of the confining ISM which we now calculate. These calculations are
mainly concerned with the prototype Seyfert~2, NGC~1068, but they are easily applied to any other
Seyfert. In this analysis we use the solution for the size of a radio lobe as given by
\citeN{bicknell97a} (see also \citeN{begelman96a}). Take $x_{\rm h}$ to be the size of the lobe (from core
to head), $x_0$ an arbitrary scale length, ambient density $\rho = \rho_0 (x/x_0)^{-\delta}$,
$F_{\rm E}$ the jet energy flux, $\zeta$ the ratio of lobe pressure near the head to average lobe
pressure and $t$ the time, then
\begin{equation}
x_h = x_0 \: 
\xi^{1/(5-\delta)} 
\label{e:x_h}
\end{equation}
where
\begin{equation}
\xi = \frac {(5-\delta)^3 \, \zeta^2}{18\pi(8-\delta)} \: 
\left( \frac { F_{\rm E} t^3}{\rho_0 x_0^5} \right) 
= 0.26 \, \frac {(5-\delta)^3}{(8-\delta)} \: 
\frac {F_{\rm E,43} t_6^3}{n_0 (x_0/{\rm kpc})^5}
\label{e:xi}
\end{equation}
where $10^{43} \, F_{\rm E,43} \> \rm ergs \> s^{-1}$ is the jet energy flux,  $t_6 \> \rm Myr$ is
the age, $n_0 \> \rm cm^{-3}$ is the ambient hydrogen density at a radius of $x_0$,  and, following
\citeN{begelman96a} we have taken $\zeta=2$. For a given size and age the implied density is
\begin{equation}
n_0 = \frac {(5-\delta)^3 \, \zeta^2}{18\pi(8-\delta)} \: 
\left( \frac { F_{\rm E} t^3}{\rho_0 x_0^5} \right) 
= 0.26 \, \frac {(5-\delta)^3}{(8-\delta)} \: 
\frac {F_{\rm E,43} t_6^3}{n_0 (x_0/{\rm kpc})^5} \: 
\left( \frac {x_{\rm h}}{x_0} \right)^{-(5-\delta)}
\end{equation}
For NGC~1068, the projected size is 700~pc and assuming that this is close to the actual size, we
obtain Hydrogen densities of 3.1, 1.3 and 0.4 $\times F_{\rm E,43} t_6^3 \: \rm cm^{-3}$ for
$\delta=0,1$ and 2 respectively.  These densities are typical of those inferred for the disk ISM in
the inner regions of spiral  galaxies (e.g.. \citeN{wevers84a}, \citeN{bosma81a}). That is, for any
reasonable density profile, energy fluxes in the range of $10^{43} -  10^{44} \: \rm ergs \> s^{-1}$
and dynamical ages $\sim 10^6 \: \rm yrs$, we have an implied range of densities $\sim 1-10 \: \rm
cm^{-3}$. This density is typical of what one would infer for most Seyferts and is in principle
observable from the diffuse emission surrounding the lobes.
\citeN{cecil90a} have determined densities $\ga 10^{2.5} \: \rm cm^{-3}$ from the [SII] lines
emitted by ionized gas in NGC~1068. If these measurements represented an average density of the ISM
in NGC~1068, then the implied jet energy flux would be substantial ($\sim 10^{45} \: \rm ergs \>
s^{-1}$). However, the ionized gas is likely to be clumpy and if the gas is shocked as we suggest,
then the [SII] density is likely to be about a factor of $50-100$ higher than the pre-shock density,
depending upon the magnetic parameter $Bn^{-1/2}$. This puts the pre-shock density
into the range suggested by the above dynamical calculation. Thus, the existing ground-based
measurements of the [SII] densities in NGC~1068 are consistent with our model but we cannot claim that they
constitute definitive support for it. Nevertheless, this analysis and further calculations given below
indicate that HST measurements of densities in this and other Seyfert galaxies are certainly of great
interest. Mapping of the Faraday rotation structure may also lead to further insights.

\newpage

\section{Implications of the Energy Budget}

There have been numerous observations of radio emission in Seyferts which indicate collimated
outflow. However, only a few observations have the required sensitivity and resolution
to actually detect jets. In the small number of cases that have been observed with sufficient
resolution, e.g.. MKN~3 
\cite{kukula93a}, MKN~6 \shortcite{kukula96a}, MKN~78 (\shortciteN{pedlar89a}, \citeN{whittle97a}),
NGC~1068 \cite{gallimore96a,gallimore96b} and NGC~4151 \cite{pedlar93a} the jets appear to be 
initially well-collimated and supersonic and then to disrupt and to become subsonic. In no cases do
there appear lobes with bright hotspots as in classical FR2 radio galaxies. However, nearer to the
core, jet knots occur, reminiscent of structures associated with shocks in supersonic jets. In
NGC~1068, at least one of the radio knots near the core appears associated with an oblique shock
where the jet is deflected by a narrow-line cloud. The northern NGC~1068 lobe may well be associated with the
end of a subsonic jet and simulations relating to this situation would be of interest. Thus, dynamically
Seyfert jets resemble FR1 jets which appear to be initially supersonic and relativistic and to then
undergo a transition to turbulent transonic flow. At this transition FR1 jets are mildly relativistic
with $\beta=v/c \approx 0.6-0.7$ \cite{bicknell94a,bicknell95a}. The question then arises, are
Seyfert jets also mildly relativistic in the region $10-100 \> \rm pc$ from the core?

Using the data in the above-cited papers it is possible to begin to address this question. In the
following we assume that the [OIII] luminosity is the result of interaction of each jet with the
ISM, i.e. the luminosity of the [OIII]$\lambda 5007$ line associated with one jet (i.e. one side
of the NLR) is given by
\begin{equation} 
L({\rm [OIII]}) \approx 0.02 \, (1-f_{\rm ad}) \, F_{\rm E}
\end{equation} 
where $f_{\rm ad} \sim 0.5$ is the adiabatic factor mentioned above, the factor
of 0.02 is the approximate fraction of the total emission line and continuum luminosity emitted in
the $\lambda5007$ line (cf. \citeN{bicknell97a}) and, as above, $F_{\rm E}$ is the jet energy flux . 
We use the energy flux inferred from this equation together with the [OIII]$\lambda5007$ 
luminosities on the jet side to estimate the jet $\beta$ using 
\begin{eqnarray}
F_{\rm E} &=& 4 p \Gamma^2 \beta c \, A_{\rm jet} \, 
\left[ 1 + \frac {\Gamma-1}{\Gamma} \chi \right]  \\
&=& 3.6 \times 10^{42} \> 
\left[ \frac {p}{10^{-8} \> \rm dyn \> cm^{-2}} \right] \> 
\left[ \frac {r_{\rm jet}}{10 \> \rm pc} \right]^2 \>
\Gamma^2 \beta \> \left[ 1 + \frac {\Gamma-1}{\Gamma} \chi \right]
\label{e:energy_flux}
\end{eqnarray}
where $p$ is the jet pressure, $\beta = \hbox{jet velocity} /c $, $\Gamma$ is the jet Lorentz factor,
$A_{\rm jet}$ is the jet cross-sectional area,  and $\chi = \rho_{\rm cold} c^2 / 4p$ the ratio of cold matter
energy density to relativistically dominated enthalpy \cite{bicknell94a}. In order to obtain indicative
estimates of jet velocity, we put $p$ equal to the minimum pressure inferred from the radio images, taking
the lower and upper cutoff Lorentz factors to be 10 and $10^4$ respectively\footnote{The minimum pressure is
relatively insensitive to these assumptions. The use of the minimum pressure here is reasonable since, if the
plasma is near equipartition, the total contribution of enthalpy and Poynting flux to the jet energy flux is
similar to the contribution implied by equation \ref{e:energy_flux} with 
$p=p_{\rm min}$.} We also take $\chi=1$, since for an initially relativistic jet making the transition to
subsonic flow, $\chi \sim 1$ \cite{bicknell96a}. However, the estimated jet velocity is fairly insensitive to
values of $\chi$ near one. 

The indicative estimates of jet velocity and Lorentz factor implied by this procedure as well as the values of
parameters used in the calculation are given in Table~1. The column labeled $f(\hbox{[OIII]})$ in the table,
is the  fraction of the {\em total} [OIII] luminosity attributed to the jet. In all but galaxy the inferred
jet velocities are of order $0.6 - 0.8 \> c$. These calculations based on the energy budget alone, therefore
could be taken to suggest that jet velocities in Seyferts are relativistic on the 10s of parsecs scale and
that they are subsequently  strongly affected by entrainment. Nevertheless, we discuss significant
difficulties with such a proposition in the following section. These are principally based upon analysis of
NGC~1068 radio and emission line data which suggest a much larger than minimum pressure in the jet and a
larger momentum flux than could be provided by a relativistic jet. If this is the case for all of the jets
that we have considered, then the real velocities are much lower.

The exceptional case (Markarian 78) in Table~1 is of interest. Because of the distance of this galaxy, it is
only possible to estimate the minimum pressure at $\sim 270 \> \rm pc$ from the core and our corresponding
estimate of the velocity is $0.05 \> c$, indicating, on the basis of the above assumptions, that the jet has
decelerated to a subsonic velocity by this stage. The jet appears to be deflected by an [OIII]-emitting cloud
at this distance from the nucleus
\cite{whittle97a} but there is no indication of a radio knot, indicating that the flow is subsonic,
consistent with a smaller velocity. This deduction of deceleration is still valid if the pressure is
dominated to the same extent by thermal plasma. 

As intimated above, energy considerations provide only a partial indication of jet velocities. Therefore, in
the next section we consider the constraints placed on jet velocities by observations of jet-cloud
interactions. This effectively brings the jet {\em momentum} budget into consideration.

\newpage

\section{The Momentum Budget from Jet-Cloud Interactions}

\subsection{General considerations}

In a jet-cloud interaction as schematically indicated in Figure~\ref{f:jet-cloud}, a supersonic jet is
deflected  by a cloud producing an oblique shock in the jet. The post-shock pressure pressure drives a shock
into the cloud which, if the cloud is dense enough, is fully radiative. The situation depicted in
Figure~{f:jet-cloud} is an oversimplification since the surface of the cloud will not remain flat but will
be "gouged out" by the jet and also become filamentary -- both as a result of the action of the
Kelvin-Helmholtz instability. This will especially be the case if the post-shock flow is subsonic. In
essence the force applied by the jet to the cloud is a fraction of the jet momentum flux and the estimates
of shock luminosity presented below based on this idealized model probably underestimate this fraction.
Nevertheless, the various parametric estimates we make below are probably indicative and motivate further
numerical work on this topic.

Given our idealized model, let us take $p_{\rm sh}$ to be the post-oblique shock pressure in the jet and
$\rho_{\rm cl}$ to be the density of the cloud. The radiative shock driven into the cloud has a velocity,
$V_{\rm sh} \approx (4 p_{\rm sh} / 3\rho_{cl})^{1/2}$ which together with the density and shock area
determines the line fluxes from the cloud. Since $p_{\rm sh} \propto
\rho_{\rm jet} V_{\rm jet}^2$ then the force exerted by the shock on the cloud,
$p_{\rm sh} A_{\rm sh}$ (where $A_{\rm sh} \> \rm cm^2$ is the shock area) is proportional to the jet {\em
momentum} flux. We estimate the various coefficients of proportionality as follows.

We parameterize the oblique shock by the pre-shock Mach number, $M_{\rm jet}$, of the incident jet and the
deflection angle,
$\Delta \theta$, of the jet, the latter being constrained by observations. We take $\theta_1$ and $\theta_2
= \theta_1 + \Delta \theta$ to be the usual angles between the pre- and post-shock flow and the shock
normal and $M_{x}=M_{\rm jet} \cos \theta_1$ to be the normal
component of Mach number. Using the Rankine-Hugoniot conditions for a non-relativistic jet, one can readily
show that 
\begin{equation}
\tan \Delta \theta = \frac {2(M_{x}^2 -1)\tan \theta_1}{M_{x}^2\left[(\gamma-1 +  (\gamma+1) \tan^2
\theta_1
\right]+2}
\end{equation}
where $\gamma=5/3$ is the ratio of specific heats.
Given $M_{\rm jet}$ and $\Delta \theta$ this equation can easily be solved numerically for $\theta_1$ and then
all of the other shock parameters, in particular,
\begin{equation}
\frac {p_{\rm sh}}{\rho_{\rm jet} V_{\rm jet}^2} = \frac {1}{\gamma M_{\rm jet}^2} + \frac {2}{\gamma+1} 
\left[ \cos^2 \theta_1 - \frac {1}{M_{\rm jet}^2} \right]
\end{equation}

The luminosity from shock excited emission lines is proportional to the shock area and we therefore need to
relate the shock area to the area of the jet. As we show in Figure~\ref{f:jet-cloud} this is determined by the
addition of $x_1$, the projection of the shock onto the cloud and $x_2$, the  distance traveled by the
jet before a sound wave emitted from the edge of the jet intercepts the cloud. The total distance is given by 
\begin{equation}
x_1+x_2 = D_{\rm jet} \sec \theta_1 \left(\sin \theta_2 + M_2 \cos \theta_2 \right),
\end{equation}
where $M_2$ is the post-shock Mach number, readily calculated form the Rankine-Hugoniot equations. The ratio of
cloud shock area to jet cross-sectional area is given by $A_{\rm sh} /A_{\rm jet} \approx (x_1+x_2)/D_{\rm
jet}$. Expressing the jet momentum flux, approximately in terms
of the energy flux by
\begin{equation}
\rho_{\rm jet} V_{\rm jet}^2 A_{\rm jet} = 
\frac {2 F_{\rm E}}{V_{\rm jet}} \left[1 + \frac{2}{(\gamma-1)M_{\rm jet}^2} \right]^{-1},
\end{equation}
we have for the force exerted by the shock:
\begin{equation}
p_{\rm sh} A_{\rm sh} \approx \frac {2F_{\rm E}}{V_{\rm jet}} 
\> f(M_{\rm jet}, \Delta \theta) 
\end{equation}
where
\begin{equation}
f (M_{\rm jet}, \Delta \theta) =
\frac {p_{\rm sh} }{\rho_{\rm jet} V_{\rm jet}^2} \, 
\left[ 1 + \frac {2}{\gamma-1} \frac {1}{M_{\rm jet}^2} \right]^{-1} 
\sec \theta_1 \left[ \sin \theta_2 + M_2 \cos \theta_2
\right]
\label{e:f_def}
\end{equation}
and the jet velocity is given by:
\begin{equation}
\beta_{\rm jet} = \frac {V_{\rm jet}}{c} \approx \frac {2F_{\rm E}}{p_{\rm sh} A_{\rm sh} c} \>
f(M_{\rm jet}, \Delta \theta).
\label{e:v_jet}
\end{equation}

The factor $f(M_{\rm jet},\Delta \theta)$ is easily estimated from the Rankine-Hugoniot conditions and a plot
of this parameter as a function of Mach number for different deflections (suggested by the projected
deflection at NLR-C) is given in  Figure~\ref{f:rh-plot}. Interestingly, the variation of $f(M_{\rm
jet},\Delta\theta)$ with Mach number is minor and there is not a great variation with deflection angle. The
physical parameter
$2F_{\rm E}/(p_{\rm sh} A_{\rm sh} c)$ can be determined from the energy flux required to power the optical
emission excited by the lobe expansion and the parameters of the radiative shock in the dense cloud as shown
below in the specific case of NGC~1068.

In Figure~\ref{f:rh-plot} we also show the ratio of post-shock to pre-shock pressure as a function of
Mach number. This is used in the following discussion of NGC~1068.

\subsection{Application to NGC~1068}

Using Merlin and VLA images together with an HST~FOC [OIII]$\lambda5007$ image of NGC~1068, 
\citeN{gallimore96b} have shown that a significant number of emission line clouds lie adjacent to 
the Northern jet in NGC~1068, in particular clouds NLR-C, NLR-D and NLR-F. The cloud NLR-G lies along
the  projected direction of the jet which may be too faint at this position to show a knot. The
projected areas of these clouds are all of order $10^{40} \> \rm cm^2$.  These emission line clouds
constitute an excellent example of where jet-cloud interactions are responsible for the excitation of
at least part of the narrow line region and the following analysis aims to determines parameters for the
radiative shocks which we assume exists in these clouds. In particular we are interested in the parameters
of the cloud NLR-C near which the radio jet has a prominent radio knot (knot~C). The jet also suffers a
clear deflection
$\approx 23^\circ$  at this location. Our interpretation of the associated jet and narrow-line cloud
morphology is that the deflection of the jet by NLR-C leads to the situation analyzed above, namely the
production of an oblique jet shock and the driving of a radiative shock into the cloud by the associated
pressure.

For shocks in the velocity range of $500-1000 \> \rm km \> s^{-1}$ the H$\alpha$ luminosity is given in
terms of the Hydrogen density $n_H \> \rm cm^{-3}$, the shock area $A_{\rm sh} \> \rm cm^2$ and the shock
velocity, $V_3$ thousand $\rm km \> s^{-1}$, by
\begin{equation}
L(H\alpha) = 5.3\times 10^{-3} n_H \,  A_{\rm sh} \, V_3^{2.41} \> \rm ergs \> s^{-1}
\label{e:h_alpha}
\end{equation}
The pressure driving the shock, $p_{\rm sh}$ is given in terms of the cloud density, $\rho_{\rm cl}$ and
the shock velocity, $V_{\rm sh}$, by
$p_{\rm sh} \approx \rho_{\rm cl} V_{\rm sh}^2$ so that the shock force is
\begin{equation}
p_{\rm sh} A_{\rm sh} \approx \rho_{\rm sh} V_{\rm sh}^2 \approx 4.5 \times 10^{-6} \> V_3^{-0.41} \,
L(H\alpha)
\label{e:shock_force}
\end{equation}
Here we use the observed velocity dispersion $\sim 1000 \> \rm km \> s^{-1}$ in the NLR of NGC~1068 as an
estimate of the shock velocity. 

The H$\alpha$ flux used in the above expression was estimated as follows: The HST H$\alpha$ filter 
transmits (redshifted) [NII]$\lambda6548$ and [NII]$\lambda6583$ and we took the transmission curve of this
filter into account to obtain an estimated H$\alpha$ flux of $F({\rm H}\alpha +[NII]) / (1+0.74R)$ where $R$
is the ratio of [NII]$\lambda6583$ to H$\alpha$. The shock models in the velocity range $500 - 1000 \:
\> \rm km \> s^{-1}$ imply that $R\approx 0.75$ so that we have used this value. Table~1 contains the
measured H$\alpha$+[NII] and [OIII] fluxes in the brightest regions of the  narrow line clouds as well as
the [OIII]:H$\alpha$ ratio. In principle one could use the velocity dependence of this ratio to estimate the
shock velocity. However, the [OIII] flux is metallicity-dependent. Moreover, the high measured values of
[OIII]:H$\alpha$ suggest to us that the observed [OIII] line flux may be dominated by the shock precursor
region. In table~3, therefore, we give estimates of number density, shock pressure and shock force for two
shock velocities, $500$ and $1000 \> \rm km \> s^{-1}$. The first two quantities are for a nominal shock
area of $10^{40} \> \rm cm^2$. As equation~(\ref{e:h_alpha}) implies the number density estimate depends
fairly strongly on the assumed shock velocity but is clearly in the vicinity of a few hundred $\rm cm^{-3}$
-- typical of molecular cloud densities. The shock pressure is much less sensitive (see
equation~(\ref{e:shock_force}) and is in the vicinity of 
$\hbox{a few} \times 10^{-6} \> \rm dynes \> cm^{-2}$ for shock areas $\sim 10^{40} \> \rm cm^{-2}$. The
shock force, of course, has the same velocity dependence as the pressure and is independent of the shock
area. This is the parameter of primary importance here since it is the shock parameter which enters into the
jet velocity estimate (see equation~(\ref{e:v_jet})). 

A clear prediction from these estimates is that densities measured from the [SII] lines should be in the
vicinity of $10^{3-4} \> \rm cm^{-3}$ because of the density of the [SII] emitting region relative to the
pre-shock density referred to earlier.

Given the parameters in tables 1 and 2, the velocity of the jet near NLR-C is estimated to be
$$
\beta_{\rm jet} \approx \frac {2 \times 10^{44}}{6 \times 10^{34} \: c} \times f (M_{\rm jet}, \Delta
\theta) \approx 0.1 \times f (M_{\rm jet}, \Delta \theta) 
$$
For a plausible range of deflections $\approx 10^\circ-30^\circ$ relevant to this particular jet-cloud
interaction,
$f (M_{\rm jet}, \Delta \theta) \sim 0.4 - 0.8$ (see Figure~\ref{f:rh-plot}) so that the jet velocity $\sim
0.04 - 0.08 \> c$. This velocity could be significantly {\em increased}  if we
have underestimated the jet energy flux required to power the NLR or if we have overestimated the force of the
radiative shock responsible for the luminosity of NLR-C or if we have underestimated the fraction of jet
momentum flux that drives the radiative shock.  If, for example, the value of the parameter $f(M_{\rm jet},
\Delta \theta)$ used above were a factor of two higher than we have  estimated here, then the jet
velocity would be a factor of two higher.  Thus a jet velocity,
$v_1 \approx 0.2-0.3 \, c$, before knot~C would not be out of the question. In view of this factor of
two uncertainty in the estimation of physical parameters further numerical simulations of this
process would be useful as would more detailed spectral observations of the NLR of NGC~1068. The
latter would assist us to better estimate the parameters of the radiative shocks involved. Despite these
reservations, this calculation demonstrates how the combination of total narrow-line region luminosities and
individual cloud luminosities associated with jet-cloud interactions can be used to derive jet velocities.
It will be interesting to see similar calculations applied to a sample of Seyferts in which the individual
uncertainties in various parameters would statistically average out. We also add that it would be unlikely
that we have estimated quantities so poorly that the above velocity is in error by an order of magnitude and
that the jet velocity is in fact close to $c$ and that the Lorentz factor is large.

Another important feature arising from this analysis is that the shock driving pressure is larger, by about
a factor of 40, than the minimum pressure inferred from the radio image and this factor is so large that
even given the uncertainties of the above analysis, it is significant.  This cannot be an effect of
resolution since the shocked region of the jet adjacent to the cloud should be extended by 2-3 jet diameters
in the jet direction (the observed elongation based upon the 20~cm. image is about 2 jet diameters) so that
the entire post-shock region should be resolved. Moreover, departures from minimum conditions of this amount
are unlikely. Rather, this is another indication that there is an additional component of the jet pressure.
We can now consider the implications of this by returning to the energy flux. Constraints derived from this
can be used to estimate the jet Mach number, as follows. For a nonrelativistic jet the energy flux can be
expressed in the form:
\begin{equation}
F_{\rm E} \approx \frac{\gamma}{\gamma-1} \, p V_{\rm jet} A_{\rm jet} \>
\left[ 1 + \frac {\gamma-1}{2} M_{\rm jet}^2 \right],
\end{equation}
where $\gamma$ is the ratio of specific heats. For a  pressure, $p = 10^{-6} p_{-6} \> \rm dyn \> cm^{-2}$ 
$V_{\rm jet} \approx 0.06 \, c$, $D_{\rm jet} \approx 18 \> \rm pc$ and $\gamma=5/3$,
\begin{equation} 
F_{\rm E} \approx 1 \times 10^{43} \> p_{-6} \left( 1 + \frac {M_{\rm jet}^2}{3}  \right).
\end{equation} 
The pressure here is the {\em pre-shock} pressure, for which we have no direct estimate. However, given that
the  {\em post-shock} pressure $\approx 6 \times 10^{-6} \left( A_{\rm sh} /10^{40} \rm cm^2 \right)^{-1}$
and that the jet energy flux, $F_{\rm E} \approx 10^{44} \> \rm ergs \> s^{-1}$ the ratio of post-shock to
pre-shock pressures, $p_2/p_1$ is given by
\begin{equation}
\frac {p_2}{p_1} \approx C \, \left( 1 + \frac{M_{\rm jet}^2}{3} \right) \: 
\left( \frac {A_{\rm sh}}{10^{40} \rm cm^2} \right)^{-1},
\label{e:p21}
\end{equation}
where $C\approx 0.6$.
The area of the radio knot adjacent to NLR-C ($\approx 5 \times 10^{39} \> \rm cm^2$) indicates that the
actual area of the shock may be less than the nominal shock area ($10^{40} \> \rm cm^2$). Therefore, in the
right panel of Figure~\ref{f:rh-plot} we have superposed plots of equation~(\ref{e:p21}),for $C=0.6$ and
$1.2$, on the plots of the theoretical shock pressure ratios. Allowing $C$ to vary by a
factor of 2 also allows for errors in the estimates other quantities such as the velocity and energy flux
which are not known to better than a factor of 2. 

It is evident from the intersection of equation~(\ref{e:p21}) with the calculated shock pressure ratios that
a low, but supersonic jet Mach number $\sim 1-\hbox{a few}$ is favored. This is consistent with the
morphology of this jet: The occurrence of knots, which it is natural to associate with shocks, is
indicative of supersonic flow; however the jet spreads at the rate that we associate with turbulent
transonic flow. Moreover, it is not as well-collimated as we would expect of a highly
supersonic jet.

Given the Mach number and velocity, the sound speed and hence the temperature of the jet can be estimated,
with the result that $T=(\mu m_{\rm p}c^2/(\gamma k) M_{\rm jet}^{-2} \beta_{\rm jet}^2 = 4.1 \times 10^{12}
M_{\rm jet}^{-2} \beta_{\rm jet}^2 \> \rm K$. For $\beta_{\rm jet}\approx 0.06$, $T\approx 1.5 \times
10^{10} M_{\rm jet}^{-2} \> \rm K$ and if $M_{\rm jet} = 3$, for example, $T=1.6 \times 10^9 \> \rm K$. Thus
the electrons are mildly relativistic and the protons are subrelativistic, assuming that they are
at the same temperature. Thus, although the particle energies are not ultrarelativistic, they are extreme,
and must be related to the plasma environment near the black hole. 

The mass flux, $\dot M$,  in the jet is also of interest. This is given by
\begin{equation}
\dot M = \frac{2F_{\rm E}}{V^2} \, \left[ 1 + \frac {2}{(\gamma -1)M_{\rm jet}^2} \right]^{-1} = 
1 \> M_{\odot} \> {\rm yr}^{-1} \> \left( \frac {F_{\rm E}} {10^{44} \> \rm ergs \> s^{-1}} \right) \>
\left[ 1 + \frac {3}{M_{\rm jet}^2} \right]^{-1}.
\end{equation} 
For values of the jet Mach number, $M_{\rm jet} = 1$, 2 and 3, the implied mass fluxes are 0.3, 0.6 and 
$0.7
\> M_{\odot} \> yr^{-1}$, respectively. These estimates are all about an order of magnitude larger than the
mass accretion rate rate, $\sim 0.05 \> M_\odot \> \rm yr^{-1}$, onto
the black hole estimated from the bolometric luminosity $\sim 3 \times 10^{44} \> \rm ergs \> s^{-1}$
(\citeN{gallimore96a} corrected for our assumed distance of 15~Mpc) and assuming an efficiency factor of
0.1. This accretion rate is close to the estimate $0.04
\> M_\odot \> \rm yr^{-1}$ of Maloney (private communication) derived from the properties of the maser
emission. The discrepancy between the jet mass flux and the accretion rate is a clear indication that most
of the jet mass flux is the result of entrainment and given the substantial amount of dense matter in the
circumnuclear environment of NGC~1068, this is not surprising. This raises the question as to whether the
{\em initial} velocity of the jet could be relativistic. To answer this we consider the implications of
momentum and energy conservation in a relativistic entraining flow. In view of our above deductions
concerning the composition of the jet plasma, we assume that the jet is thermally dominated and that
initially, the energy and momentum fluxes are dominated by the rest-mass inertia. Hence the initial energy
and momentum fluxes ($F_{E,1}$ and
$F_{p,1}$ respectively) are given by
\begin{eqnarray}
F_{E,1} &\approx& (\Gamma_1-1) \, \dot M_1 c^2  \\
F_{p,1} &\approx& \Gamma_1 \, \dot M_1 c \beta_1
\end{eqnarray}
where $\beta_1$ is the initial value of $v/c$, $\Gamma_1$ is the initial bulk Lorentz factor and $\dot
M_1$ is the initial flux of rest mass. The energy flux is conserved, irrespective of entrainment. If the
ambient pressure gradient does not have a major effect on the momentum flux (as is the case when the Mach
number is high), then it also is approximately conserved. Thus the ratio of these two
quantities is also conserved. When the flow becomes subrelativistic the ratio of energy to momentum flux is
$V_{\rm jet}/2(1+3/M_{\rm jet}^2)$ so that
\begin{equation}
\frac {(\Gamma_1 -1)c}{\Gamma_1 \beta_1 } \approx 
\frac {1}{2} \, v \, \left( 1 + \frac {3}{M^2} \right) 
\label{e:e_p_ratio}
\end{equation}
and the initial velocity is determined by the velocity and Mach number at knot~C by
\begin{equation}
\, \frac {\Gamma_1-1}{\Gamma_1 \beta_1} \approx 
\beta_{\rm jet} \frac {1+3/M_{\rm jet}^2}{2}   
\label{e:beta_ep}
\end{equation}
The nonrelativistic limit of this equation is 
\begin{equation}
\beta_1 \approx \beta_{\rm jet} \left( 1 + \frac {3}{M_{\rm jet}^2} \right)
\end{equation}
These equations imply that for a transonic Mach number $\sim 1-2$ near knot C the initial velocity of the jet
is about 2-4 times higher, i.e. about $0.1 - 0.3 c$, given our above estimates. As we have outlined above, it
is not unreasonable to envisage ``factor of order unity'' changes in the parameters of the radiative shock,
jet energy flux and the fraction of the jet momentum flux driving the radiative shock that have gone into the
estimate of the jet velocity at knot~C.  Moreover, if the background pressure gradient increases the
momentum flux, in the transonic regime of the jet, then equation~(\ref{e:beta_ep}) underestimates the jet
velocity on the tens of parsecs scale, with the result that the initial jet velocity could be higher. (This
can be seen more clearly from equation~(\ref{e:e_p_ratio}) with $\approx$ replaced by $\ga$.) Thus it is
reasonable to conclude that the initial jet velocity  in NGC~1068 is mildly relativistic and this is what
we expect from a flow ejected from, say within 10 gravitational radii of a black hole. However, it is
unlikely that the Lorentz factor is of order a few--10, as in radio galaxies and quasars.

The following considerations of the mass flux enable us to relate the jet properties to that of the corona
above the nuclear accretion disk. The initial mass flux, $\dot M \approx F_{\rm E}
c^{-2} (\Gamma_1 -1)^{-1}$, i.e. 
$2F_{\rm E}E (c \beta_1)^{-2} \approx 4 \times 10^{-3} \beta^{-2} F_{\rm E,44} \> M_\odot \> \rm yr^{-1}$, in
the non-relativistic limit. If the jet is mildly relativistic (say $\beta \sim 0.5$) then the implied mass
flux $\sim 0.02 \> M_\odot \> \rm yr^{-1}$, close to, but less than, the above-estimated mass accretion rate
$\sim 0.05
\, M_\odot \> \rm yr^{-1}$. (The two are equal for $\beta_1\approx 0.3$.) We suppose that the jet originates
from the coronal region of the black hole accretion disc and that the appropriate radius is  of order 10
gravitational radii since this is where most of the accretion disc and related coronal dissipation occur. 
It is now reasonably well-established that the hard X-ray emission from Seyferts is thermally Comptonized
emission from an accretion disc corona with an electron temperature,
$T_{\rm e} \sim 50-100 \> \rm keV$ and Thomson optical depth, $\tau_T \sim 2$ 
\cite{johnson97a}. Adopting a coronal radius $r_c \sim 10 \> \hbox{gravitational radii}$, then the electron
density in the corona, 
\begin{equation}
n_{\rm c} \sim 1.0 \times 10^{11} \, \tau_{\rm T} (r_c/10 r_g)^{-1} M_7^{-1} \> \rm cm^{-3}
\end{equation}
where $r_{\rm g}=GM/c^2$ is the gravitational radius for a black hole of mass $M$, and
$10^7 M_7 M_\odot$ is the black hole mass. Assuming that some of the corona is ejected in the form of a
mildly relativistic jet of density $n_1$, radius $r_1 \sim 10 r_g$, and a Lorentz factor $\Gamma_1$ then the
corresponding mass flux, 
\begin{eqnarray}
\dot M_1 &\approx& \pi r_1^2 n_1 \Gamma_1 \beta_1 c \nonumber \\
&\approx& \frac {\pi \mu m_{\rm p} G^2}{c^3} \> M^2 \left( \frac {r_1}{10 r_g} \right)^2 
\> n_1 \Gamma_1 \beta_1  \nonumber \\
& \approx & 3.5 \times 10^{-2} \>  \tau_{\rm T}  
\left( \frac {r_1}{10r_{\rm g}} \right)^2 \,
\left( \frac {r_{\rm c} }{10 r_g} \right)^{-1} \, M_7 \,
\left( \frac {n_1}{n_{\rm c}} \right) \, \Gamma_1 \beta_1 \>
M_\odot \> \rm yr^{-1}
\end{eqnarray}
Given that $M_7 \approx 3$ for NGC~1068, assuming that $\tau \approx 2$, that the jet velocity is mildly
relativistic (again for argument, say $\beta_1 \sim 0.5$) and also assuming that the jet density
$n_1 \sim n_{\rm c}$ then the initial jet mass flux $\sim 0.1 \> M_\odot \> \rm yr^{-1}$. This estimate
is close to  the mass flux estimated from the jet dynamics and any discrepancy could be attributed to the 
decrease of density from the corona to where the jet terminal velocity is established. 
Assuming that the jet and coronal radii are similar (i.e. $r_1\sim r_{\rm c}$), then our estimate
is not very sensitive to our assumptions concerning these parameters. This good order of magnitude agreement
motivates closer examination of the relationship between discs, their coronae and the jets ejected from them.

The other issue to address here is the confinement of a jet with a pressure $\ga 10^{-6} \> \rm dynes
\> cm^{-2}$. This is too large a pressure to be provided by a hot interstellar medium; a number density
$\ga 700 \> \rm cm^{-3}$ would be required and the cooling from this would be catastrophic. The other
possibility, suggested by the observations of \citeN{cecil90a}, is that the jet is inertially confined by a
dense cool outflowing wind. If the wind has a density $\sim 100 \> \rm cm^{-3}$, the overpressured jet
drives shocks into it with a velocity
\begin{equation} 
V{\rm sh} = 650 \> {\rm km \> s^{-1}} \> \left( \frac {P}{10^{-6} \> \rm dynes \> cm^{-2}}
\right)^{1/2}
\, \left( \frac {n}{100 \> \rm cm^{-3}} \right)^{1/2}
\end{equation} 
There is evidence for such shocks from the observations of \citeN{axon97a} who find
spectral evidence for line splitting of approximately $1000 \> \rm km \> s^{-1}$ in the gas coincident with
the jet (in projection). The presence of the coronal line of [FeVII]$\lambda 3769$ in their spectra is
indicative of high excitation. Using the data of \citeN{cecil90a} which indicate a velocity $\sim 1000 \>
\rm km \> s^{-1}$ and an opening cone angle for the wind $\sim 40^\circ$, the mass flux is $\dot
M_{\rm wind}
\approx 0.8 \> M_\odot \> \rm yr^{-1}$. (We have evaluated the cross-sectional area at 24~pc, the
location of NLR-C.) Again, it is interesting that this mass flux is of order the mass accretion rate. The
energy transported by such a wind $\sim 3 \times 10^{41} \> \rm ergs \> s^{-1}$ and therefore the wind is
not competitive with the jet in exciting the narrow-line region. 

Finally, we also note here that the jet velocity implied by a conventional (radio galaxy) value of the
parameter
$\kappa_\nu$ would be unreasonably small. The value of $\kappa_{1.5}$ implied by our calculated jet energy
flux
$\sim 10^{44} \> \rm ergs \> s^{-1}$ and the
\citeN{wilson82a} radio flux density, $3.8 \> \rm Jy$, for the linear  part of the source is $\kappa_{1.5}
\sim 5.5 \times 10^{-15} \> \rm Hz^{-1}$.  If
$\kappa_{1.5}$ were a factor of $10^3-10^4$ higher, say $\sim 10^{-12} - 10^{-11} \> \rm Hz^{-1}$, the jet
energy flux would be a factor of
$10^3-10^4$ lower and the jet velocity, implied by equation~(\ref{e:v_jet}), would be in the range of
$3-30 \> \rm km \> s^{-1}$. Therefore, our interpretation of the jet and associated emission-line
morphology, provides additional support for a much lower value of $\kappa_\nu$ in NGC~1068 and consequently
for much greater energy input into the NLR than previously supposed. The only way in which this
conclusion could be in error would be if the cloud NLR-C were photoionized and its excitation had nothing to
do with the deflection of the jet.

\newpage

\section{Discussion}

In this paper we have addressed the physics which is necessary in order to support the view that the NLR
emission from Seyfert galaxies be powered, at least in part, by  the energy and momentum associated with the
radio jets. One requires jet energy fluxes $\sim 10^{43-44} \> \rm ergs \> s^{-1}$ and a much smaller ratio
of radio power to jet energy flux than is normally invoked for radio galaxies. A lower value of this ratio is
suggested by the smaller ages of Seyferts compared to radio galaxies. However, we also require that either
the magnetic fields in the lobes of Seyferts be sub-equipartition or more likely, that the fraction by energy
of relativistic plasma in the lobes be much smaller than in radio galaxies. This raises the question as to
where the composition of the jets is established. Are the jets initially relativistic in composition and
subsequently diluted by entrainment of thermal plasma from the ISM or are they initially 
thermally dominated outflows with a small proportion
of embedded relativistic gas? If we adopt the first point of view and assume that all AGN jets have initially
relativistic velocities and that their internal energies are initially dominated by relativistic plasma, then
we infer mildly relativistic velocities in four Seyfert jets mostly within $100
\> \rm pc$ of the core. On the other hand, in NGC~1068, additional information provided by a jet-cloud
interaction near the core, suggests a lower velocity $\sim 0.04 - 0.0.08 \> c$ 24~pc from the core and, in
order to maintain the same energy flux, a substantial contribution to the jet pressure from thermal gas is
required. As we noted above, the velocity estimate from the parameters of the jet-cloud interaction also
support a significantly higher jet energy flux in the northern NGC~1068 jet than previously supposed and
therefore support our suggestion of a much lower value of the parameter $\kappa_\nu$ in Seyfert galaxies. A
counter to this argument would involve the proposition that NLR-C is, in fact, photoionized, and that the
deflection of the jet at this point has nothing to do with its excitation. This could be resolved by
detailed HST optical and UV spectra of this cloud which could establish whether it is photoionized or
shocked (see, for example, \citeN{dopita97a}).

At present the notion that Seyfert jets may not be initially ultrarelativistic in composition, is only based
upon the analysis of data on one galaxy, NGC~1068. However, the alternative, that Seyfert jets are dominated
by thermal plasma from the outset, is probably more sustainable as a general proposition.  If the internal
energy of Seyfert jets were relativistically dominated, then one would have to explain why a $\sim 10^{44}
\> \rm ergs \> s^{-1}$ jet of relativistic plasma produced so little radio emission on kpc scales. Dilution by
substantial entrainment of thermal gas and a related energy loss by the relativistic component as it mixes
with the thermal plasma may go part way to explaining the low level of extended radio emission. However, a
model of this sort may meet substantial physical difficulties. It is far more straightforward to invoke a
substantial thermal component from the outset. Nevertheless, whilst we have
shown, on the basis of our interpretation that the NGC~1068 jet is probably not dominated by
ultrarelativistic particles, the inferred temperature of the NGC~1068 jet, at 24~pc from the core, $\sim
10^{9}-10^{10} \> \rm K$ is substantial and must be related to the environment of the black hole where it
originated. As we have shown, a high pressure jet must interact substantially with its surroundings via
shocks and the high excitation line emission coincident with the jet, discovered by \shortciteN{axon97a}, is
good evidence for this. The mass flux in the wind, $\sim 1 \> M_\odot \> \rm yr^{-1}$ is similar to that in
the jet at that distance from the core, and is similar to the mass accretion rate onto the black hole. 

Work by \citeN{colbert97a} supports the view that Seyfert radio plasma is, at least on large scales, dominated
by thermal gas. He finds that, in Seyferts with large-scale outflows, the overwhelming contribution to the
lobe pressure is provided by soft X-ray emitting gas, the radio emitting plasma contributing about 1\%
consistent with our requirements on the energy flux. Dominance of the lobe pressure by ($T\sim 10^7 \> K$),
thermal gas has another physical consequence which enhances the self-consistency of our
model: The isobaric cooling time for shocked gas,
$t_{\rm cool}
\approx 3.6
\, p_{-9}^{-1} T_7^{2.45}
\> \rm Myr$ (where $10^{-9} p_{-9} \>
\rm dyn \> cm^{-2}$ is the pressure and $10^7 \, T_7 \> \rm K$ is the temperature) indicates lobe cooling
time scales $\sim \hbox{a few} \rm \> Myr$. This provides an explanation for why the dynamical ages of
Seyferts appear to be $\sim \hbox {a few} \> \rm Myr$ (corresponding to characteristic expansion velocities
$\sim 500 \> \rm km \> s^{-1}$ and sizes $\sim \rm kpc$). Once the gas in the lobe cools, the lobe collapses
and then builds up again over this timescale. Obviously, the gas in the initial region of the jet needs to
cool from $\sim 10^9 \> \rm K$ both adiabatically and through mixing in order to reach $\sim 10^7 \> \rm K$.

The paper by \citeN{nelson95a} is also of interest in this context. They have shown that when the {\em total}
powers of Seyfert and radio galaxies are plotted against bulge magnitude, the radio galaxies lie well above
the sequence defined by the Seyferts. However, restriction to the {\em core} powers of radio galaxies sees
the radio galaxies continuing the Seyfert sequence. That is, when the radio powers of Seyfert and radio
galaxies are compared on similar scales, they form part of the same sequence. This suggests that entrainment
on the kpc scale may have some influence on the level of kpc-scale radio emission. However, the Seyferts do
have, on average, core powers which are lower than those of the radio galaxies consistent with them being
weaker radio sources from the outset.

Since the mass flux estimated for the jet at 24~pc from the core exceeds the accretion rate into the black
hole by about an order of magnitude, it is likely that the jet mass flux at this point is the result of
entrainment. Our estimate the jet velocity near the black hole is subject to
the uncertainties in the estimates of jet velocity and Mach number near knot~C. However, it is not
hard to justify a mildly relativistic velocity. The jet velocity near the black hole is also related to the
initial mass flux. If the initial velocity is mildly relativistic then the mass flux is similar to but less
than the accretion rate. Moreover, the mass flux is consistent with a typical Seyfert coronal density $\sim
10^{11} \> \rm cm^{-3}$ being ejected at a mildly relativistic velocity. Thus consideration
of the NLR excitation has led us to a fundamental linkage between jet properties and the coronal properties
of accretion discs.

The remaining point we wish to make is that, for Seyfert jets dominated by thermal plasma, one
generally expects a large amount of internal Faraday rotation. For example at knot C in NGC~1068, we
calculate a rotation measure of $20,000 \> \rm rad \> m^{-2}$ for an equipartition field $\sim 10^{-3} \>
\rm G$. Unfortunately internal jet depolarization is difficult to disentangle from  depolarization by the
NLR.

 We are grateful to Prof. Mark Whittle and Prof. Andrew Wilson for
many helpful discussions over the course of this research, and for access to their data on Markarian 78
before publication. We are also grateful to Dr. S. Baum for helpful discussions on NGC~1068 in the early
stages of this work, to Dr. Greg Madejski for providing us with the extremely useful preprint of
\citeN{johnson97a} and to Drs. Zdenka Kuncic and Phil Maloney for useful discussions .

\newpage

\bibliography{apjmnemonic,gvbrefs}
\bibliographystyle{apjv2}

\newpage

\begin{sideways}

{
\begin{tabular}{l c c c c c c  l}
\multicolumn{8}{c}{\bf Table 1: Jet velocities inferred from [OIII] and radio data} \\
\\
\hline
\multicolumn{1}{c}{Galaxy \&}           &
\multicolumn{1}{c}{Distance to}        &
\multicolumn{1}{c}{Diameter}         &
\multicolumn{1}{c}{$p_{\rm min}$}        &
\multicolumn{1}{c}{$f([OIII])$}          &
\multicolumn{1}{c}{$L([OIII])$}          &
\multicolumn{1}{c}{$\beta$}              &
\multicolumn{1}{c}{Radio reference}      \\
\multicolumn{1}{c}{Component}            &
\multicolumn{1}{c}{nucleus (pc)}         &
\multicolumn{1}{c}{(pc)} &
\multicolumn{1}{c}{$\rm dyn \> cm^{-2}$} &
&
\multicolumn{1}{c}{$\rm ergs \: s^{-1}$}  &
\multicolumn{1}{c}{} &
\multicolumn{1}{c}{Optical reference} \\
\hline
Mkn 3 E    & 69  & 41    & $2.6 \times 10^{-8}$   & 0.5     & $4.8 \times 10^{41}$     &  0.63 &\citeN{kukula93a} \\  
Mkn 3 W    & 141 & 36    & $2.0 \times 10^{-8}$   & 0.5     & $4.8 \times 10^{41}$     &  0.73 &  \citeN{koski78a}  \\
\hline
Mkn 6 4    & 127 & 51    & $5.8 \times 10^{-8}$   & 0.5     & $1.7 \times 10^{42}$     &  0.63  & \citeN{kukula96a} \\
           &     &       &       &         &         &       &  \citeN{koski78a} \\
\hline
Mkn 78 W   & 271 & 180   & $5.7 \times 10^{-8}$   & 1.0$^1$ & $8.7 \times 10^{41}$    &  0.05  & \citeN{whittle97a} \\
           &     &       &       &         &         &       &  \citeN{whittle97a} \\
\hline
NGC 1068  C & 24  & 18    & $1.4 \times 10^{-7}$    & 1.0$^2$ & $1.0 \times 10^{42}$    &  0.77  & \citeN{gallimore96a} \\
            &     &       &       &         &         &       &  \shortciteN{storchi95a}   \\
\hline
NGC 4151 C1& 114 & 37    & $9.5 \times 10^{-9}$  & 0.5     & $2.2 \times 10^{41}$       &  0.71  & \citeN{pedlar93a} \\   
NGC 4151 C2& 56  & 20    & $1.4  \times 10^{-8}$   & 0.5     & $2.2 \times 10^{41}$     &   0.73 & \citeN{anderson70a}\\  
NGC 4151 C5& 114 & 23    & $1.4 \times 10^{-8}$   & 0.5     & $2.2 \times 10^{41}$      &   0.82 &  \\
\hline
\\
\multicolumn{8}{l}{$^1$ [OIII] luminosity of western region of source.}\\
\multicolumn{8}{l}{$^2$ The entire [OIII] luminosity  of NGC 1068 is assigned to the northern region because of the one-sided
ionization cone.}\\
\end{tabular}




}
\end{sideways}

\newpage

\begin{sideways}


\begin{tabular}{c c c c c c c}
 \\
\multicolumn{7}{c}{\bf Table 2: Observed H$\alpha$ and [OIII]$\lambda5007$ luminosities} \\
\\
\hline
\multicolumn{1}{c}{Cloud} &
\multicolumn{1}{c}{$F(H\alpha + [NII])$} &
\multicolumn{1}{c}{$ F(H\alpha)$} & 
\multicolumn{1}{c}{$\log L(H\alpha)$} & 
\multicolumn{1}{c}{$ F([OIII])$} &
\multicolumn{1}{c}{$ \log L([OIII])$} & 
\multicolumn{1}{c}{[OIII]/H$\alpha$} \\
&
\multicolumn{2}{c}{$ \rm ergs \> cm^{-2} \> s^{-1}$} &
\multicolumn{1}{c}{$\rm ergs \> s^{-1}$} &
\multicolumn{1}{c}{$\rm ergs \> cm^{-2} \> s^{-1}$} &
\multicolumn{1}{c}{$ \rm ergs \> s^{-1}$} &
\multicolumn{1}{c}{ } \\

\hline
NLR-C & $8.50 \times 10^{-13}$ & $5.47\times 10^{-13}$ & 40.14 & $2.03 \times 10^{-12}$ & 40.71 &
3.7 \\
NLR-D & $4.01 \times 10^{-13}$ & $2.58\times 10^{-13}$ & 39.81 & $9.00\times 10^{-13}$  & 40.35 &
3.5\\
NLR-F & $4.40 \times 10^{-13}$ & $2.83\times 10^{-13}$ & 39.85 & $1.50 \times 10^{-12}$ & 40.57 &
5.2 \\
NLR-G & $2.59 \times 10^{-13}$ & $1.67\times 10^{-13}$ & 39.62 & $9.99 \times 10^{-13}$ & 40.40 &
6.0 \\
\hline
\end{tabular}


\end{sideways}



\vskip 1 cm

\begin{tabular}{c c c c c}
 \\
\multicolumn{5}{c}{\bf Table 3: Estimated number densities and shock pressures} \\
\\
\hline
Cloud     & $V_{\rm sh}$ & $n_H(A_{\rm sh}/10^{40} \: \rm cm^2)$  & $p_{\rm sh} A_{\rm sh}$ 
& $p_{\rm sh}(A_{\rm sh}/10^{40} \: \rm cm^2)$       \\
         & $\rm km \> s^{-1}$ & $ \rm cm^{-3}$   & dynes & $\rm dynes \> cm^{-2}$ \\
\hline \\
NLR-C     & 500 & 1500 & $8.2 \times 10^{34}$  & $8.2 \times 10^{-6}$  \\
NLR-D     & 500 & 690   & $3.8 \times 10^{34}$ & $3.8 \times 10^{-6}$\\
NLR-F     & 500 & 750   & $4.2 \times 10^{34}$ & $4.2 \times 10^{-6}$ \\
NLR-G     & 500 & 440   & $2.5 \times 10^{34}$ & $2.5 \times 10^{-6}$ \\
\\
\hline
\\
NLR-C     & 1000 & 280  & $6.2 \times 10^{34}$ & $6.2 \times 10^{-6}$ \\
NLR-D     & 1000 & 130   & $2.9 \times 10^{34}$ & $2.9 \times 10^{-6}$ \\
NLR-F     & 1000 & 140   & $3.2 \times 10^{34}$ & $3.2 \times 10^{-6}$\\
NLR-G     & 1000 & 80   & $1.9 \times 10^{34}$ & $1.9 \times 10^{-6}$ \\
\\
\hline

\end{tabular}


\newpage

\begin{center}
{\Large \bf Figure captions}
\end{center}

\begin{figure}
\centering \leavevmode
\includegraphics[bb=18 154 578 693,width=\textwidth]{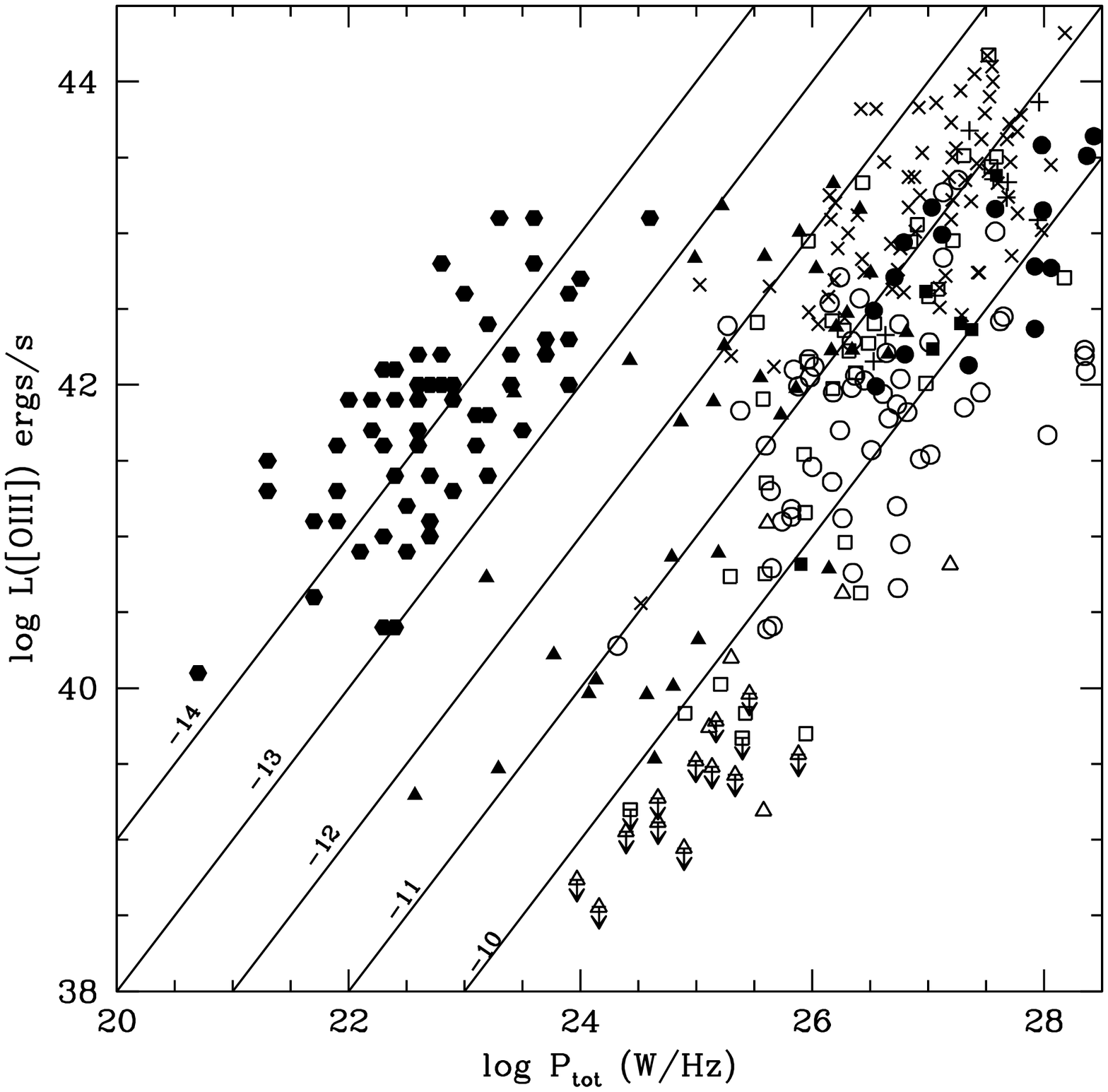}
\caption{}
\label{f:oiii}
\end{figure}

\begin{figure}
\centering \leavevmode
\includegraphics[bb=18 154 578 693,width=\textwidth]{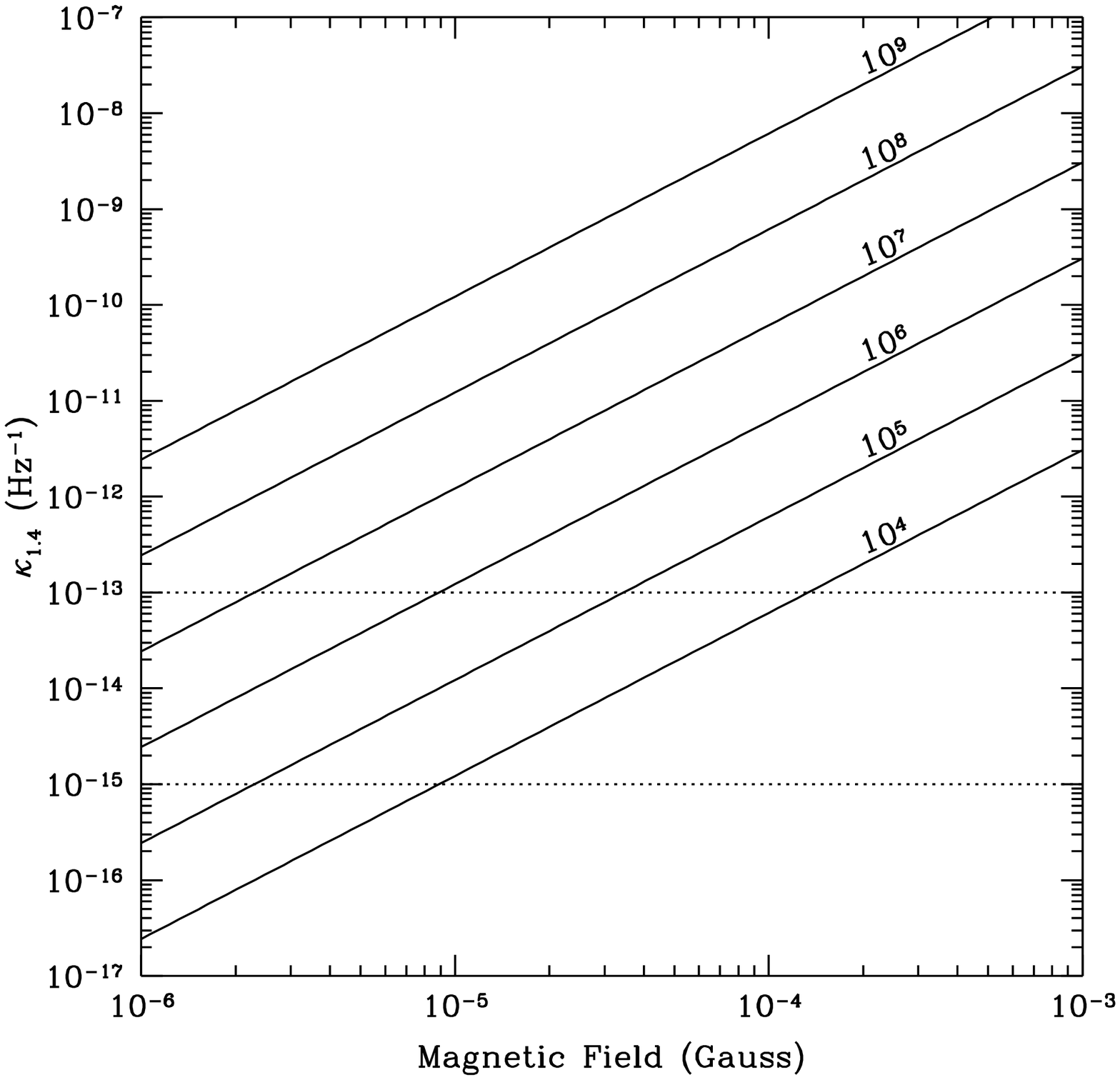}
\caption{}
\label{f:kappa_nu}
\end{figure}

\begin{figure}
\centering \leavevmode
\includegraphics[width=\textwidth]{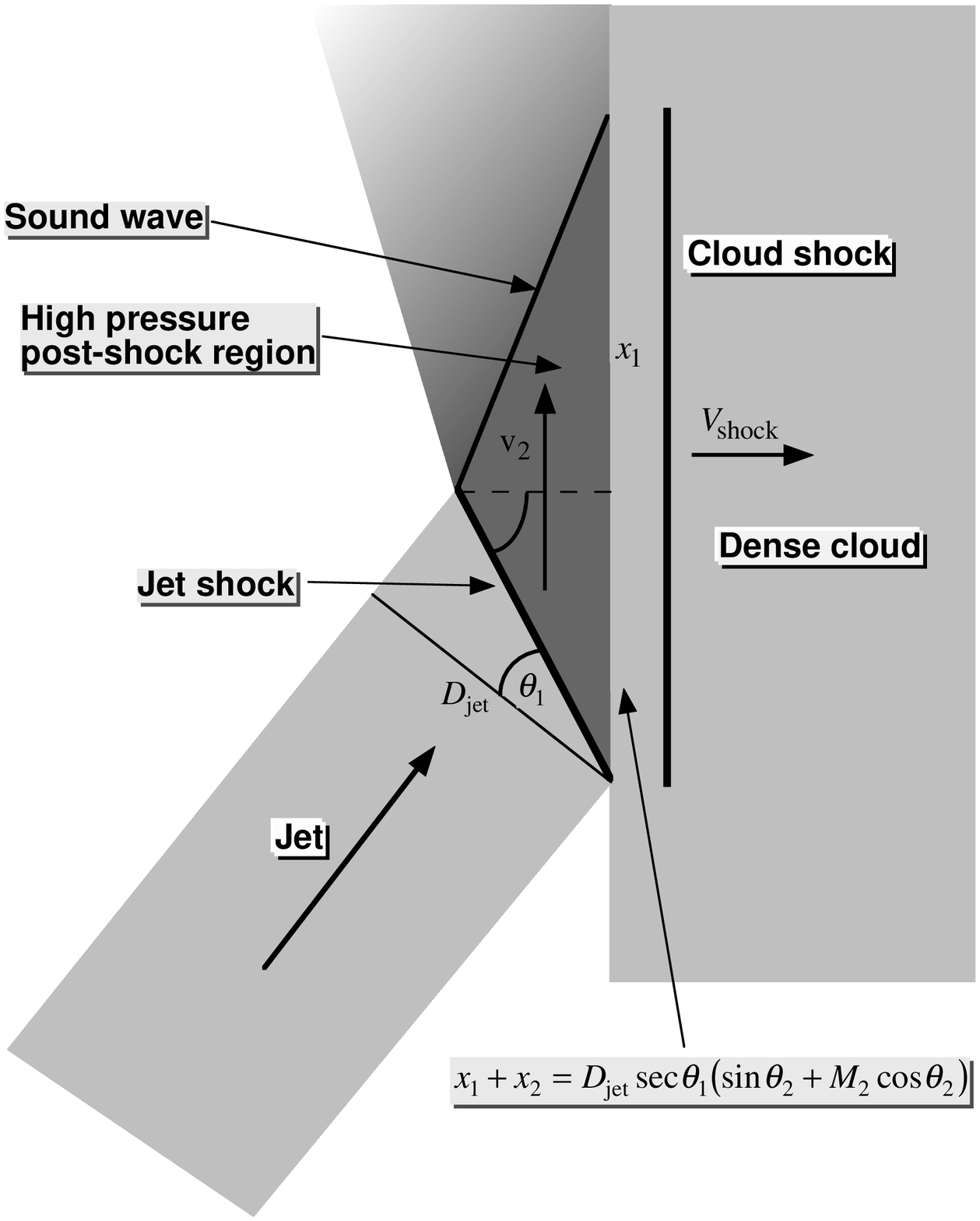}
\caption{}
\label{f:jet-cloud}
\end{figure}

\begin{figure}
\centering \leavevmode
\includegraphics[width=\textwidth]{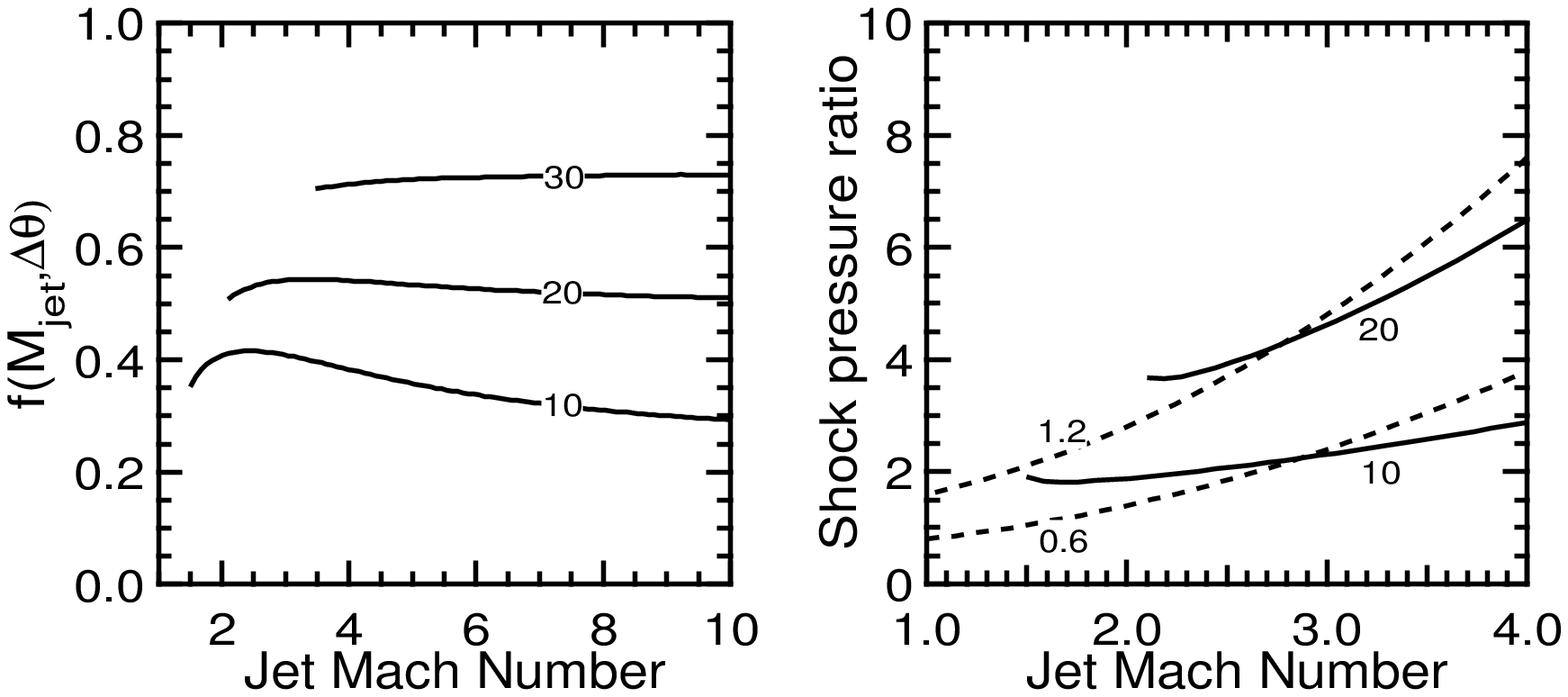}
\caption{}
\label{f:rh-plot}
\end{figure}

\begin{itemize}

\item[\bf Figure \ref{f:oiii}:] Predicted $[OIII]$ luminosity as a function of 1.4 GHz  radio power
for values of $\log \kappa_{1.4} = -10, -11, -12, -13$  and -14 overlaid on data for radio-loud
objects and Seyfert galaxies. Data are from \citeN{gelderman96a} (GW), \citeN{tadhunter93} and
\citeN{morganti93} (collectively referred to as TM) and \citeN{whittle85}. Legend: Filled circles -- CSS
sources from GW; filled squares -- CSS sources from TM; open circles -- GW FR2 radio galaxies; open squares
-- TM FR2 radio galaxies;  diagonal crosses -- GW QSOs; plus signs: TM compact flat spectrum sources; open
triangles: TM FR1 radio galaxies; filled hexagons: Seyfert galaxies from \citeN{whittle85}. Upper
limits are indicated in the usual way.

\item[\bf Figure \ref{f:kappa_nu}:] The ratio $\kappa_{1.4}$ of radio power at 1.4~GHz to jet energy
flux for different values of the age parameter $\tau = f_{\rm e} f_{\rm ad} t$. A spectral index of
0.7 and a lower  cutoff, $\gamma_0 = 1.0$ have been assumed. However, the value of $\kappa_\nu$ is
not very sensitive to these parameters.

\item[\bf Figure \ref{f:jet-cloud}:] This figure defines the parameters of a jet-cloud interaction. The extent
of the high pressure post-jet-shock region is determined by the sound wave which propagates towards the cloud
from the edge of the jet. This high pressure region drives a radiative shock into the cloud.

\item[\bf Figure \ref{f:rh-plot}:] The left panel shows the factor $f(M_{\rm jet},\theta_1)$ defined by
equation~(\ref{e:f_def}) which parameterizes the force of the post jet-shock region ($p_{\rm sh} A_{\rm sh}$)
in units of $F_{\rm E}/V_{\rm jet}$ for jet deflections ($10^\circ$, $20^\circ$ and $30^\circ$) as
indicated. The right panel shows the pressure ratio (solid lines) $p_2/p_1$ as a function of pre-shock jet
Mach number for the given deflections. Superimposed on these curves are dashed curves ($p_2/p_1 = C(1+M_{\rm
jet}^2/3)$, $C=0.6$ and 1.2) representing the pressure ratio implied by the radiative shock analysis and the
jet energy budget for NGC~1068. Both panels are for a non-relativistic jet with the ratio of specific heats
$\gamma=5/3$.

\end{itemize}

\newpage

\newpage

\newpage

\newpage

\end{document}